\documentclass[11pt]{article}
\usepackage{amsmath}
\begin{document}

\thispagestyle{empty}

\begin{flushright} LPTENS-05/27 \end{flushright}

\vskip 0.5cm

\begin{center}{\LARGE { GAUGE THEORIES AND  
\vskip 0.4cm
 NON-COMMUTATIVE GEOMETRY}}

\end{center}

\vskip1cm
\begin{center}
{\bf{E.G. Floratos}}

\vskip0.2cm
Physics Dept. University of Athens, Greece\\
mflorato@mail.uoa.gr

\vskip0.5cm
{\bf{J. Iliopoulos}}
\vskip0.2cm

Laboratoire de Physique Th\'eorique CNRS-ENS\\ 
24 rue Lhomond, F-75231 Paris Cedex 05, France\\
ilio@lpt.ens.fr
\vskip1.0cm

{\bf ABSTRACT}

\end{center}

\vskip 0.5cm

It is shown that a $d$-dimensional classical $SU(N)$ Yang-Mills theory can be formulated in a
$d+2$-dimensional space, with the extra two dimensions forming a surface with
non-commutative geometry. In this paper we present an explicit proof for the
case of the torus and the sphere. 

\bigskip

\newpage


Quantum field theories on a non-commutative space-time have been studied for a
long time \cite{wess1}. Not surprisingly, the first one to suggest such a
formulation was W. Heisenberg, as early as 1930 \cite{wess1}. His motivation
was to introduce an ultraviolet cut-off to cure the short-distance
singularities which appeared to plague all attempts to construct a
relativistic quantum field theory. Such field theories on a space with non-commuting coordinates have indeed been 
formulated \cite{snyder}, although the motivation did not turn out to be a
very valid one. With the development  of the renormalisation programme, the
problem of ultraviolet divergences took a completely different turn. The
geometry of physical space may still produce an ultraviolet cut-off, but its
presence is not relevant for the calculation of physical processes among
elementary particles at foreseeable energies.

However at the same time, a new motivation for studying theories in a
non-commutative space appeared, although only recently it was fully
appreciated. In 1930 L.D. Landau \cite{landau} solved the problem of the
motion of an electron in an external constant magnetic field and, besides
computing the energy levels, the so-called ``Landau levels'', he showed that
the components of the velocity operator of the electron do not
commute. Following Heisenberg's suggestion, R. Peierls \cite{peierls} showed
that, at least  the lowest Landau level, can be obtained by using this
space non-commutativity. Since the presence of non-vanishing magnetic-type
external fields is a common feature in many modern supergravity and string
models, the study of field theories formulated on spaces with non-commutative
geometry \cite{connes} has become quite fashionable \cite{genrev}. A new
element was added a few years ago with the work of N. Seiberg and E. Witten
\cite{sw} who showed the existence of a map between  gauge theories formulated
in  spaces with commuting and non-commuting coordinates. In this paper we want
to present a different but related result which is inspired by the behaviour
of $SU(N)$ Yang-Mills theories for large $N$. An earlier version of this work
has been presented in \cite{fi1}. We shall state and, to a certain extend,
prove, the following statement:   

\vskip0.5cm

{\it Statement:} Given an $SU(N)$ Yang-Mills theory  in a $d-$dimensional space
with potentials

\begin{equation}
\label{gaugepot}
 A_{\mu}(x)~=~A_{\mu}^a (x)~t_a 
\end{equation}
where $t_a$ are the standard $SU(N)$ matrices, there exists a reformulation in
which  
the gauge fields and the gauge potentials become:  

\begin{equation}
\label{newlimits}
A_{\mu}(x) \rightarrow {\cal A}_{\mu}(x,z_1,z_2)~~~~~~~ F_{\mu \nu}(x) \rightarrow {\cal F}_{\mu \nu}(x,z_1,z_2)
\end{equation}
where ${\cal A}$ and ${\cal F}$ are fields in a $(d+2)-$dimensional space,
greek indices still run from 0 to $d-1$ and $z_1$ and $z_2$ are 
local coordinates on a  two-dimensional surface endowed with non-commutative
geometry. They will be shown to satisfy the commutation relation

\begin{equation}
\label{fuzsphcom}
[z_1,z_2]=\frac{2i}{N}
\end{equation}

Notice that the parameter which determines the non-commutativity is related to the rank of the group. The commutators in  the original $SU(N)$ Yang-Mills theory are
replaced by the Moyal brackets \cite{moyal}, \cite{fairlie} with respect to
the non-commuting coordinates.

\begin{equation}
\begin{split}
\label{commoy}
[A_{\mu}(x), A_{\nu}(x)] & \rightarrow \{{\cal A}_{\mu}(x,z_1,z_2), {\cal
    A}_{\nu}(x,z_1,z_2)\}_{Moyal}\\
[A_{\mu}(x),\Omega(x)] & \rightarrow \{{\cal A}_{\mu}(x,z_1,z_2), 
 {\it   \Omega} (x,z_1,z_2)\}_{Moyal}
\end{split}
\end{equation}
where $\Omega $ is the function of the gauge transformation and $\{,\}_{Moyal}$
denotes the Moyal bracket with respect to the two operators $z_1$ and $z_2$.
The trace over the group indices in the original Yang-Mills action becomes a
two dimensional integral over the surface:

\begin{equation} 
\label{action}
\int d^4x~ Tr\left( F_{\mu \nu}(x)F^{\mu \nu}(x)\right)~~\rightarrow~~ \int d^4xdz_1dz_2~{\cal F}_{\mu \nu}(x,z_1,z_2)*{\cal F}^{\mu \nu}(x,z_1,z_2)
\end{equation}

The *-product will be defined later. When $N$ goes to infinity, the two $z$'s
commute and the *-product reduces to the ordinary product.

\vskip 0.5 cm 

In what follows we shall give a
partial proof of this statement.
\vskip 0.5 cm



Let us start by recalling a well-known algebraic result: The Lie algebra of
the group $SU(N)$, at the limit when $N$ goes to infinity, with the generators appropriately 
rescaled, becomes the algebra of the area preserving diffeomorphisms of a
surface. There exist explicit proofs of this theorem for the case of the
sphere and the torus \cite{hoppe}, \cite{fi2}, \cite{zachos1}. This implies a
corresponding field theoretic result: A classical $SU(N)$ Yang-Mills theory on
a $d$-dimensional space at an appropriate large $N$ limit is equivalent to a
field theory on  $d+2$ dimensions with the matrix commutators replaced by
Poisson brackets with respect to the two new coordinates \cite{fi3},
\cite{zachos2}, \cite{sakita}. This is just the large $N$ limit of the
relations (\ref{gaugepot})-(\ref{action}) we want to prove, in other words we
want to establish in this paper the equivalence between Yang-Mills
theories and field theories on surfaces to any order in
$1/N$. 

\vskip 0.5 cm

In order to be specific, let us consider first the case of the torus. One way to prove the algebraic result is to isolate, inside $SU(N)$, a quantum $U(1)\times U(1)$ group. It is convenient to distinguish the case with $N$ odd or even. For odd $N$ we define two $N\times N$ unitary, unimodular matrices, (Weyl matrices), by:

\begin{equation}
\label{weylmatr}
g~=~\left(\begin{array}{ccccc}
1 & 0 & 0 & ... & 0 \\
0 & \omega & 0 & ... & 0 \\
0 & 0 & \omega^2 & ... & 0 \\
. & . & . & .~~ & 0 \\
. & . & . & ~.~ & 0 \\
. & . & . & ~~. & 0 \\
0 & 0 & 0 & ... & \omega^{N-1} 
\end{array} \right)~~~;~~
h~=~\left(\begin{array}{ccccc}
0 & 1 & 0 & ... & 0 \\
0 & 0 & 1 & ... & 0 \\
. & . & . & .~~ & . \\
. & . & . & ~.~ & . \\
. & . & . & ~~. & . \\
0 & 0 & 0 & ... & 1 \\
1 & 0 & 0 & ... & 0
\end{array} \right)
\end{equation}
where $\omega$ is an $N$th root of unity $\omega=exp(4\pi i/N)$. They satisfy
quantum group commutation 
relations:

\begin{equation}
\label{ghrel}
g^N~=~h^N~=~1~~~~~;~~~hg=\omega gh
\end{equation}

The important point is that integer mod$N$ powers of $h$ and $g$ can be used to construct the generators of $SU(N)$:

\begin{equation}
\label{torsun}
S_{m_{1}, m_{2}}~=~\omega^{m_{1} m_{2}/2}g^{m_1}h^{m_2}
\end{equation}

Under hermitian conjugation we obtain $S_{m_{1}, m_{2}}^{\dagger}=S_{-m_{1}, -m_{2}}$ and the algebra closes with the help of

\begin{equation}
\label{torsunalg}
S_{\bf m} S_{\bf n} =\omega^{{\bf n}\times {\bf m}/2}S_{{\bf m}+{\bf n}}~~~~~~~~~[S_{\bf m} ,S_{\bf n} ]=2isin\{ (2\pi/N)({\bf m}\times {\bf n})\} S_{{\bf m}+{\bf n}}
\end{equation} 

The normalisation of the generators can be obtained from the trace relations

\begin{equation}
\begin{split}
TrS_{m_{1},m_{2}}=0 ~~~~~{\rm except ~for}~~~~~m_1 =m_2 =0 ~modN \\
TrS_{\bf m}S_{\bf n}~=~N\delta_{{\bf m}+{\bf n},{\bf 0}}~~~~~~~~~~~~~~~~~~~~~~~
\label{suntraces}
\end{split} 
\end{equation}
where we have used the notation ${\bf n}=(n_{1}, n_{2})$ and ${\bf n}\times
{\bf m}=n_{1}m_{2}-m_{1}n_{2}$

One can show \cite{zachos2} that, as $N$ goes to infinity, the $SU(N)$ algebra (\ref{torsunalg}), 
with generators rescaled by a factor proportional to $N$,  becomes isomorphic 
 to that of the area preserving diffeomorphisms of a two dimensional torus. A similar construction can be made for $N$ even. 

The connection between $SU(N)$ at $N \rightarrow \infty$ and $[SDiff(T^2)]$ can be made explicit by choosing a pair of variables forming local symplectic coordinates on the torus, for example, the angles $z_1$ and $z_2$ of the two circles, and expanding all functions on the torus on the basis of the eigenfunctions of the laplacian:

\begin{equation}
\label{torsol}
h_{n_{1},n_{2}}=exp(in_{1}z_{1}+2\pi in_{2}z_{2}),~~~n_{1},n_{2} \in Z\!\!\!Z
\end{equation}

Here we are interested in the fuzzy torus, so we endow $z_1$ and $z_2$ with the commutation relations of the Heisenberg algebra (\ref{fuzsphcom}). If we define the corresponding group elements $h$ and $g$, by:

\begin{equation}
\label{heisgr}
h=e^{i z_1}~~~~;~~~~g=e^{-2i\pi z_2}
\end{equation}
we shall prove that the two commutation relations (\ref{fuzsphcom}) and
(\ref{ghrel}) are equivalent for the set of group elements $h^{n_1}$ and
$g^{n_2}$ with $n_1$, $n_2$ integers mod$N$. Notice that the later imply the
algebra of $SU(N)$. The generators of the Heisenberg algebra $z_i$ and the group elements $h$ and $g$ of (\ref{heisgr}) are infinite dimensional operators, but we can represent the $SU(N)$ algebra by the finite dimensional ones (\ref{weylmatr}) and (\ref{torsun}). They form a discreet subgroup of the Heisenberg
group and they have been used to construct quantum mechanics on a discreet
phase space \cite{fn}. In this case, we can define two new operators
$\hat {q}$ and $\hat {p}$, the first being the position operator on the
discreet configuration space and the second its finite Fourier
transform. They can be represented by $N\times N$ matrices, but, obviously, they  do not satisfy anymore the Heisenberg
algebra \cite{fl}. 

\vskip 0.5cm        

The proof of the equivalence between (\ref{fuzsphcom}) and
(\ref{ghrel}) is straightforward. 

The direct part, $i.e.$ from the algebra (\ref{fuzsphcom}) to the group (\ref{ghrel}), is a consequence of the Cambell-Hausdorff relation.

\begin{equation}
\label{ch1}
hg=e^{i(z_1-2\pi z_2)-2i\pi/N}=e^{4i\pi/N}gh
\end{equation}

The field theoretic result of (\ref{action}) follows from this part alone. 

The opposite connection, namely from (\ref{ghrel}) to (\ref{fuzsphcom}), is also true in the following sense: Let $z_1$ and $z_2$ be two operators satisfying the commutation relation

\begin{equation}
\label{op}
[z_1,z_2]=i\xi t_1(z_1,z_2)+i\xi^2t_2(z_1,z_2)+...
\end{equation}
where we have assumed an expansion in powers of the non-commutativity parameter $\xi$. $t_1$, $t_2$, ... are arbitrary operators, functions of $z_1$ and $z_2$. We can show that (\ref{ghrel}) imply that $\xi t_1=2/N$ and all $t_k=0$ for $k>1$.
The essence of the story is that any corrections
on the r.h.s. of (\ref{fuzsphcom}) will affect the quantum group commutation relations (\ref{ghrel}).  The
argument is inductive, order by order in $\xi$. Let us start with the first term and write the general form of (\ref{fuzsphcom}) as:

\begin{equation}
\label{;}
[z_1,z_2]=i\xi t_1(z_1,z_2)+O(\xi^2)
\end{equation}
 Using (\ref{;}) we compute the first order  term in the commutation relation of $h$ and $g$ given by (\ref{heisgr}). If we impose that they satisfy (\ref{ghrel}) we determine $\xi t_1(z_1,z_2)$:

\begin{equation}
\label{;;;;;;}
\xi t_1(z_1,z_2)=\frac{2}{N}
\end{equation}

We can now go back to (\ref{;}) and determine the next term in the
expansion. We write:

\begin{equation}
\label{!}
[z_1,z_2]=\frac{2i}{N}+\frac{1}{N^2}t_2(z_1,z_2)+O(\frac{1}{N^3})
\end{equation}

We look now at the commutator of $h$ and $g$ to next order and imposing always (\ref{ghrel}) we get 

\begin{equation}
\label{!!!!}
t_2(z_1,z_2)=0
\end{equation}

It is now clear how the induction works: We assume the commutator

\begin{equation}
\label{!!!!!}
[z_1,z_2]=\frac{2i}{N}+\frac{1}{N^k}t_k(z_1,z_2)+O(\frac{1}{N^{k+1}})
\end{equation}
and set the coefficient of the corresponding correction in the quantum  commutation
relation (\ref{ghrel}) equal to zero. This gives again:

\begin{equation}
\label{**}
t_k(z_1,z_2)=0
\end{equation}

\vskip 1cm

The commutation relation (\ref{fuzsphcom}) is the main step of the argument. Any
function $f$ of the $SU(N)$ generators, in particular any polynomial of the Yang-Mills
fields and their space-time derivatives, can be rewritten, using 
 (\ref{torsun}), as a function of $z_1$ and $z_2$. Since they satisfy the
quantum mechanics commutation relations (\ref{fuzsphcom}), the usual proof of
Moyal \cite{moyal} goes through and the commutator of two such functions $f$ and $g$ will
have an expansion in powers of $1/N$ of the form:

\begin{equation}
\label{moyal}
[f,g] \sim \frac{1}{N} \{f(z_1,z_2),g(z_1,z_2)\} + \frac{1}{N^2}
\left(\{\frac{\partial f}{\partial z_1}, \frac{\partial
  g}{\partial z_2}\} - \{\frac{\partial f}{\partial z_2}, \frac{\partial
  g}{\partial z_1}\} \right) +...
\end{equation}

\noindent with the Poisson brackets defined the usual way: 

\begin{equation}
\label{poisson} 
\{f,g\} = \left(\frac{\partial f}{\partial z_1} \frac{\partial
  g}{\partial z_2} - \frac{\partial f}{\partial z_2} \frac{\partial
  g}{\partial z_1}\right)  
\end{equation}

The first term in this expansion is unambiguous but the coefficients of the
higher orders depend on the particular ordering convention one may adopt. For
example, in the symmetric ordering, only odd powers of $1/N$ appear. 

For the symmetric ordering, we can introduce, formally, a *-product through:

\begin{equation}
\label{star}
f(z)*g(z)=exp(i\xi ~ \epsilon _{ij}~\partial _z^i \partial _w^j)f(z)g(w)|_{w=z}
\end{equation}

\noindent with $z=(z_1,z_2)$ and $\xi =\frac{2}{N}$. The $SU(N)$ commutators
in the Yang-Mills Lagrangian can now be replaced by the *-products on the
non-commutative surface. This equality will be exact at any given order in the
$1/N$ expansion. This completes the proof of our statement. 

\vskip 1cm

Let us turn now to the case of the sphere. One way to
prove the algebraic result \cite{fi2}, is to  start with the remark that the spherical harmonics $Y_{l,m}(\theta ,\phi)$ are
harmonic homogeneous polynomials of degree $l$ in three euclidean coordinates $x_{1}$, 
$x_{2}$, $x_{3}$:

\begin{equation}
\label{sphcoord}
x_{1}=cos\phi ~sin\theta,~~~~x_{2}=sin\phi ~sin\theta, ~~~~x_{3}=cos\theta
\end{equation}

\begin{equation}
\label{Ylm}
Y_{l,m} (\theta, \phi)=~~\sum _{i_{k}=1,2,3 \atop k=1,...,l}
\alpha_{i_{1}...i_{l}}^{(m)}~x_{i_{1}}...x_{i_{l}}
\end{equation}
where $\alpha_{i_{1}...i_{l}}^{(m)}$ is a symmetric and traceless tensor. For fixed $l$ there are 
$2l+1$ linearly independent tensors $\alpha_{i_{1}...i_{l}}^{(m)}$,
$m=-l,...,l$. 

Let us now choose, inside $SU(N)$, an $SU(2)$ subgroup by choosing three $N\times N$ hermitian 
matrices which form an $N-$dimensional irreducible representation of the Lie algebra of $SU(2)$.

\begin{equation}
\label{su2}
[S_{i},S_{j}]=i\epsilon_{ijk}S_{k}
\end{equation}

The $S$ matrices, together with the $\alpha$ tensors introduced before, can be
used to construct a basis of $N^2-1$ matrices acting on the fundamental
representation of $SU(N)$ \cite{schwinger}. 

\begin{equation}
\begin{split}
S^{(N)}_{l,m}=~~\sum _{i_{k}=1,2,3 \atop k=1,...,l}
\alpha_{i_{1}...i_{l}}^{(m)}~S_{i_{1}}...S_{i_{l}}  \\
[S^{(N)}_{l,m},~S^{(N)}_{l',m'}]=if^{(N)l'',m''}_{l,m ;~ l',m'}~S^{(N)}_{l'',m''}
\label{sunalg}
\end{split}
\end{equation}
where the $f'$s appearing in the r.h.s. of (\ref {sunalg}) are just the $SU(N)$ structure constants in a 
somehow unusual notation. 
The important, although trivial, observation is that the three $SU(2)$ generators $S_{i}$, 
rescaled by a factor proportional to $1/N$, will have well-defined limits as
$N$ goes to infinity \cite{fi3}.

\begin{equation}
\label{rescsu2gen}
S_{i}\rightarrow T_{i} = \frac {2}{N} S_{i}
\end{equation}

Indeed, all matrix elements of $T_i$ are bounded by $|(T_i)_{ab}|\leq$ 1. 
They satisfy the rescaled algebra:

\begin{equation}
\label{rescsu2}
[T_{i},T_{j}]=\frac {2i}{N} \epsilon _{ijk}T_{k}
\end{equation}
and the Casimir element

\begin{equation}
\label{casimir}
T^2=T_{1}^2+T_{2}^2+T_{3}^2=1-\frac {1}{N^2}
\end{equation}
in other words,  under the norm $\| x \| ^2 = Trx^2 $, the limits as $N$ goes to infinity of the 
generators $T_{i}$ are three objects $x_{i}$ which commute by (\ref{rescsu2}) and are constrained 
by (\ref{casimir}).

If we consider two polynomial functions $f(x_{1},x_{2},x_{3})$ and $g(x_{1},x_{2},x_{3})$ the corresponding matrix polynomials  $f(T_{1},T_{2},T_{3})$ and $g(T_{1},T_{2},T_{3})$ have commutation relations for large $N$ which follow from (\ref{rescsu2}):

\begin{equation}
\label{limpois}
\frac {N}{2i}~ [f,g] \rightarrow ~~\epsilon_{ijk} ~x_{i}~ \frac {\partial {f}}{\partial {x_{j}}} \frac {\partial {g}}{\partial {x_{k}}} 
\end{equation}

If we replace now in the $SU(N)$ basis (\ref {sunalg}) the $SU(2)$ generators $S_{i}$ by the rescaled ones $T_{i}$, we obtain a set of $N^2-1$ matrices $T^{(N)}_{l,m}$ which, according to (\ref{Ylm}), (\ref{sunalg}) and (\ref{limpois}), satisfy:

\begin{equation}
\label{limpoisY}
\frac {N}{2i}~ [T^{(N)}_{l,m},T^{(N)}_{l',m'}] \rightarrow ~\{ Y_{l,m},Y_{l',m'} \}
\end{equation}

The relation (\ref{limpoisY}) completes the algebraic part of the proof. It shows that the 
$SU(N)$ algebra, under the rescaling (\ref{rescsu2gen}), does go to that of
$[SDiff(S^2)]$. Since the classical fields of an $SU(N)$ Yang-Mills theory can
be expanded in the basis of the matrices $T^{(N)}_{l,m}$, the relation
(\ref{limpois}) proves also the field theoretical result.

\vskip 0.5cm

Here we want to argue again that the equivalence between Yang-Mills
theories and field theories on surfaces is in fact valid to any order in
$1/N$. Following the same method we used for the torus, we parametrise the $T_i$'s in terms of two operators,
$z_1$ and $z_2$. As a first step we write:

\begin{equation}
\label{fuzsph}
T_{1}=cosz_1 ~(1-z_2^2)^{\frac{1}{2}},~~~~T_{2}=sinz_1 ~(1-z_2^2)^{\frac{1}{2}}, ~~~~T_{3}=z_2
\end{equation}

A similar parametrisation has been given by T. Holstein and H. Primakoff in terms of creation and annihilation operators \cite{hp}. At the limit of $N$ $\rightarrow$ $\infty$, they become the coordinates
$\phi$ and $cos\theta$ of a unit sphere. To leading order in  $1/N$, the commutation
relations (\ref{rescsu2}) induce the commutation relation (\ref{fuzsphcom}) between
the $z_i$'s.

In higher
orders, however, the definitions (\ref{fuzsph}) must be corrected because the
operators $T_1$ and $T_2$ are no more hermitian. It turns out that  
a convenient choice is to use $T_+$ and $T_-$. We thus write:

\begin{equation}
\begin{split}
\label{fuzsphnew}
T_+ & =T_1+iT_2=e^{\frac{iz_1}{2}}~(1-z_2^2)^{\frac{1}{2}}~e^{
  \frac{iz_1}{2}}\\
T_- & =T_1-iT_2=e^{-\frac{iz_1}{2}}~(1-z_2^2)^{\frac{1}{2}}~e^{-
  \frac{iz_1}{2}}\\
T_3 & =z_2
\end{split}
\end{equation}

Here again we want to emphasise that (\ref{fuzsph}) and (\ref{fuzsphnew}) are not invertible as matrix relations and do not imply a representation of the $z$'s in terms of finite dimensional matrices. They should be understood as providing a way to express 
 the $SU(N)$ algebra in two equivalent ways:
We can start from the non-commutative coordinates of the fuzzy sphere  $z_1$
and $z_2$ which are assumed to satisfy the quantum mechanical commutation
relations (\ref{fuzsphcom}). Through (\ref{fuzsphnew}) we define
three operators $T_1$, $T_2$ and $T_3$. We can prove that they satisfy
exactly, without any higher order corrections, the $SU(2)$ relations
(\ref{rescsu2}) and (\ref{casimir}) and, consequently, they can be used as
basis for the entire $SU(N)$ algebra. 
The opposite is also true. The $SU(2)$ commutation relations (\ref{rescsu2}) imply the quantum mechanical
relation (\ref{fuzsphcom}). We can express the $SU(2)$ generators $T_i$
$i=1,2,3$, through (\ref{fuzsphnew}), in terms of two operators $z_i$ $i=1,2$. We can
again prove that they must satisfy the Heisenberg algebra (\ref{fuzsphcom}) to all orders
in $1/N$. The proof is identical to that we presented for the torus. 

\vskip 1cm

Before closing, a few remarks:

The *-product we defined in (\ref{star}) is symmetric in $z_1$ and
$z_2$. However, in order to write the Yang-Mills Lagrangian on the sphere we
must use symmetric polynomials in the $T_i$'s using the Baker-Hausdorff
*-product for $SU(2)$ \cite{ka}. The correction terms between the two expressions can be explicitly computed in any order in $1/N$. 

The proof of the equivalence between Yang Mills theories and field theories on non-commutative surfaces has been given only for the case of the torus and the sphere. In principle, however, such a formulation should be possible
for arbitrary genus surfaces, \cite{novikov}, \cite{jaffe}, \cite{ba}, although we do not
know of any explicit proof. 

It is straightforward to generalise these results and include matter fields,
provided they also belong to the adjoint representation of $SU(N)$. In
particular, the supersymmetric Yang-Mills theories have the same property. The
special case of ${\cal N}=4$ supersymmetry is of obvious interest because of
its conformal properties.

 We believe that one could also
include fields belonging to the fundamental representation of the gauge
group. In 't Hooft's limit such matter multiplets are restricted to the
edges of the diagram, so we expect in our case the generalisation to
involve open surfaces.

The equivalence between the original $d$-dimensional Yang-Mills theory and the
new one in $d+2$ dimensions holds at the classical level. For the new
formulation however, the ordinary perturbation series at the large $N$ limit is divergent. The reason is that the quadratic
part of this action does not contain derivatives with respect to
$z_1$ or $z_2$. This is not surprising because these
divergences represent the factors of $N$ in the diagrams of the
original theory which have
not been absorbed in the redefinition of the coupling constant.
However, we expect a perturbation expansion around some appropriate
non-trivial classical solution to be meaningful and to contain
interesting information concerning the strong coupling limit of the
original theory. 

A final remark: Could one have anticipated the emergence of this action in the
$1/N$ expansion? It is clear that, starting from a set of $N$ fields
$\phi^i(x)$ $i=1,...,N$, we can
always replace $\phi^i(x)$, at the limit when $N$ goes to
infinity, with $\phi (\sigma, x)$ where $0 \leq \sigma \leq 2 \pi
$. In this case the sum over i will become an integral over
$\sigma $. However, for a general interacting field theory, the $\phi^4$ term  will no more be local in
$\sigma $. So, the only surprising feature  is that,
for a Yang-Mills theory, the resulting expression is local.

\end{document}